\documentclass{article}
\usepackage{authblk}
\usepackage{mathptmx}       
\usepackage{helvet}         
\usepackage{courier}        
\usepackage{type1cm}        
\usepackage{makeidx}         
\usepackage{graphicx}        
\usepackage{multicol}        
\usepackage[bottom]{footmisc}
\usepackage{verbatim}
\usepackage{changepage}
\usepackage{rotating}
\usepackage{setspace}
\usepackage{lineno}

\date{}

\usepackage{epsfig}
\usepackage{amsmath}
\usepackage{amsmath,amssymb}
\usepackage{bbm}
\usepackage{bm}
\usepackage{amsfonts}
\usepackage{amssymb}
\usepackage{color}
\usepackage{epic}
\usepackage{tikz}
\usetikzlibrary{calc}
\usetikzlibrary{positioning,shadows.blur}
\usepackage{pifont}

\newcommand*\Laplace{\Delta}
\usepackage{pgfplots}
\usepackage{siunitx}
\usepackage[export]{adjustbox}
\usepackage{cite}
\usepackage{comment}
\usepackage{listings}
\usepackage{booktabs,siunitx}
\title{Global optimization using Gaussian Processes to estimate biological parameters from image data}

\author[1,2]{Diana Barac}
\author[1,2]{Michael D. Multerer}
\author[1,2 *]{Dagmar Iber}

\affil[1]{Department for Biosystems Science and Engineering, ETH Zurich, Mattenstrasse 26, 4058 Basel, Switzerland}
\affil[2]{Swiss Institute of Bioinformatics (SIB), Mattenstrasse 26, 4058 Basel, Switzerland}
\affil[*]{Corresponding author}
\affil[ ]{\textit {diana.barac@bsse.ethz.ch, michael.peters@bsse.ethz.ch, dagmar.iber@bsse.ethz.ch}}

\makeindex             

\definecolor{SteelBlue}{rgb}{0.27, 0.51, 0.71}
\definecolor{LightSteelBlue}{rgb}{0.6902,    0.7686 ,   0.8706}
\begin{document}
\maketitle


{\bf Declarations of interest:} None.

\clearpage

\section{Abstract}
Parameter estimation is a major challenge in computational modeling of biological processes. This is especially the case in image-based modeling where the inherently quantitative output of the model is measured against image data, which is typically noisy and non-quantitative. In addition, these models can have a high computational cost, limiting the number of feasible simulations, and therefore rendering most traditional parameter estimation methods unsuitable. In this paper, we present a pipeline that uses Gaussian process learning to estimate biological parameters from noisy, non-quantitative image data when the model has a high computational cost. This approach is first successfully tested on a parametric function with the goal of retrieving the original parameters. We then apply it to estimating parameters in a biological setting by fitting artificial in-situ hybridization (ISH) data of the developing murine limb bud. We expect that this method will be of use in a variety of modeling scenarios where quantitative data is missing and the use of standard parameter estimation approaches in biological modeling is prohibited by the computational cost of the model.

\clearpage

\section{Introduction}
Computational models of biological systems typically serve two purposes: to test the validity of underlying assumptions and to make predictions about the behaviour of the system \cite{kitano2002computational}.  Such models often contain parameters whose true values are unknown. This can be because the parameters correspond to physical constants e.g. production rates, that are difficult or expensive to measure accurately, or it can be that the parameter has no direct biological equivalent that can be determined experimentally. The behaviour of the model, and hence the predictions derived from it, can differ depending on the parameter set used. Consequently, parameter estimation is an important problem in computational modeling, see \cite{ingram2006network, geier2012analyzing, jaqaman2006linking, kreutz2009systems} and the references therein. 

Recent advances in imaging technologies, which enable ever increasing spatial and temporal resolution, have resulted in widespread use of image data in computational modeling \cite{gomez2017image}.  The unique advantage of image data such as in situ hybridization (ISH) or immunohistochemistry (IHC) data lies in is its ability to reveal the spatial and temporal distribution of a molecule of interest in an anatomically intact system \cite{megason2007imaging}. However, in most cases, this data is non-quantitative and noisy, making its integration into a quantitative computational model challenging. In addition, biological processes are generally comprised of a large number of coupled components that react in a non-linear fashion and whose behavior may vary in both space and time. This spatio-temporal behaviour can be modelled by coupled non-linear partial differential equations (PDEs), e.g. reaction-diffusion equations, whose solution usually entails a high computational cost. In this paper, we address the problem of parameter estimation for image-based modeling in which the model is computationally expensive and the data is noisy and non-quantitative. 

Parameter estimation aims to find the set of parameters for which the model best fits the available data. To this end, the first step consists of defining a suitable cost function which measures the distance between the model output and the available data, and is therefore dependent on the type of data used. When quantitative data is available, a common choice for the cost function is the sum of squared deviations (least squares) \cite{sivia2006data}. We look at the case where the model output and available data are images, thus, it is necessary to define an appropriate image metric that compares the non-quantitative data with the quantitative model output. The image Euclidean distance (applying the sum of squared deviations to image pixels) is the most commonly used image metric \cite{wang2005euclidean}. We describe a way to define the cost function, taking into account the non-quantitative nature of the image data. To this end we weaken this metric to a pseudometric by decomposing the image into areas of high and low intensity, prioritizing that the model recapitulates those rather than a set gradient, which can vary between images. 

The next step in parameter estimation is the minimization of the cost function with respect to the parameters. If the cost function has a single minimum (or the parameters can be restricted to a range such that only one minimum is present) local optimization methods can be used. These can be gradient-based or direct, see e.g. \cite{fletcher2013practical} and the references therein. However, as computational models are often complex and nonlinear, this generally leads even simple cost functions, such as least squares, to possess multiple minima. Then, it is necessary to use a global parameter estimation method as a local one will likely result in the algorithm getting trapped in a local minimum. The necessity of global estimation methods in biological modelling is discussed in \cite{mendes1998non, moles2003parameter}.

There exist several algorithms for global optimization, see \cite{horst2013handbook, moles2003parameter} and the references therein. The earliest and simplest of these methods is multi-start (running a local method repeatedly from a set of predefined initial starting points). However, this approach usually does not work for realistic applications, due to its computational inefficiency. Simulated Annealing is widely applied in global optimization problems and was shown in  \cite{mendes1998non} to give the best solution in a comparison of parameter estimation methods to estimate rate constants. Nonetheless, the authors noted the high computational cost of the algorithm. Evolutionary Computation methods (including Genetic Algorithms, Evolutionary Programming and Evolution Strategies) are widely used in global optimization \cite{goldberg1988genetic, holland1992adaptation, Fogel1966, beyer2002evolution}. Based on the concept of genetic evolution, these methods take an initial population of parameters and iteratively refine it (usually by selecting the best parameters and recombining them in some way) until an optimum parameter set is found. Hence, these methods also require a large number of individual simulations. Hybrid methods, i.e. the combination of different optimization methods, have also been used successfully in parameter estimation of biological models \cite{rodriguez2006hybrid, rodriguez2006novel, fomekong2007efficient}.

An alternative approach is to infer the posterior distribution over the parameters (the probability of the parameters given the data) instead of just retrieving a point estimate. This can be achieved using Bayesian methods \cite{wilkinson2007bayesian}. As the calculation of the likelihood function is often analytically and computationally intractable, approximate Bayesian computation (ABC) methods, which replace this calculation with a comparison between observed and simulated data, are a popular choice. ABC methods based on Markov chain Monte Carlo (MCMC) and sequential Monte Carlo (SMC) have been successfully applied to biological systems \cite{toni2009approximate, sisson2007sequential}. As these involve sampling a large number of parameters to estimate the posterior distribution, these approaches are usually only feasible for lower-dimensional problems where model simulation is computationally cheap.

As can be seen, most of the global optimization methods used in biological parameter estimation require a large number of independent model simulation steps, making them well suited for models that are computationally cheap to solve, but impractical for models whose simulation is computationally expensive. Surrogate or response surface modeling is an approach to global parameter estimation that is applied when the number of possible model evaluations is greatly limited \cite{jones1998efficient}. Response surface methods aim to estimate the response surface (the surface of the cost function as a function of the parameters) in an adaptive fashion and usually consist of two steps that are iterated \cite{jones2001taxonomy}. In the first step, the response surface is estimated using the model outputs that have been computed thus far. The second step consists of selecting a suitable next point for evaluation. This is usually based on a tradeoff between exploration (selecting a point in a region where the model has not been evaluated) and expected improvement (selecting a point that is most likely to give a lower value of the response surface). Different functions can be used to estimate the shape of the response surface from the given data. One such class is regression models, where the response surface is assumed to take a particular shape (e.g. quadratic or linear) and the surface is fitted accordingly. This works well if the assumed shape is indeed an accurate assumption of the response surface, but poorly otherwise \cite{jones2001taxonomy}. Another class is interpolating models, where the response surface is made up of basis functions based on the value of the response surface at known parameter values. Gaussian processes are one example of such a method \cite{rasmussen2004gaussian} and can be used to estimate the response surface and the uncertainty of that estimate at each point, thereby facilitating step 1 and step 2 of the response surface method. Although Gaussian processes are applied in biological modeling, their use is rather restricted to predicting the state variables of a model \cite{lawrence2007modelling, calderhead2009accelerating}, than directly estimating the cost function with respect to the parameters. In this paper, we apply Gaussian processes to estimate the cost function and thereby estimate the parameters of our model. 

Parameter dependencies are frequently present in biological models \cite{gutenkunst2007universally}. If a model parameter cannot be unambiguously estimated from the available data it is said to be non-identifiable. Determining the presence of parameter dependencies is important as model predictions may differ depending on the model parameters i.e., two sets of parameters that fit the available data equally well may give rise to different model predictions. Non-identifiable parameters can be either structurally or practically non-identifiable. A parameter is structurally non-identifiable if the model structure prevents it from being estimated unambiguously, independent of the quality of the data. A parameter is practically non-identifiable if the amount or quality of the available data (rather than the structure of the model) prevents its unambiguous estimation. Structural non-identifiability is reviewed in \cite{cobelli1980parameter} and both structural and practical non-identifiability are discussed in \cite{raue2009structural}.

For biological systems described by ordinary differential equation (ODE) models, various methods for detecting structural non-identifiability have been proposed \cite{chis2011structural}. For arbitrary models, the profile likelihood approach, a general approach to identify structural and practical non-identifiability, is introduced in \cite{raue2009structural}. The identifiability of the parameter is investigated by varying it and looking at the minimum objective function value at each point under all possible values of the remaining parameters. However, if the model is not analytically tractable and is computationally costly, calculating this profile likelihood may not be feasible. We use multivariate sensitivity analyses to determine parameter dependencies, and therefore parameter identifiability, in our models as the above approaches are not appropriate due to the models' ``black box" nature and high computational cost.

The rest of the paper is structured as follows: in the next section we describe the parameter estimation process. This is split into two parts: the first considers defining the cost function in the absence of non-quantitative data and the second describes the parameter estimation process. The following section presents the results. To test the method, we first use it with images generated from a parametric function with the goal of retrieving the original parameters. We then present the result of estimating parameters in a more complex biological setting with the aim of fitting artificial in-situ hybridization (ISH) data of the developing murine limb bud. The final section is the discussion where we consider the impact of non-identifiability, the potential limitations of the process and future work. 

\section{Method}
\subsection{Cost Function}
\begin{figure} [h]
\centering
\includegraphics[width=0.55\textwidth]{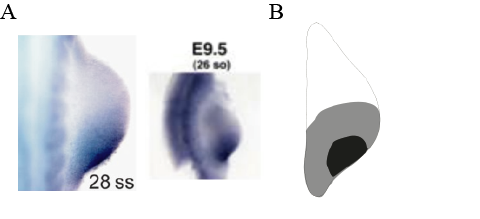}
\caption{Representation of the image data in the cost function. (A) Examples of two ISH images of a mouse limb bud at a similar developmental stage stained to reveal the expression domain of \textit{Ptc1}. The images were reproduced from \cite{zhang2017shh} (left) and \cite{zhu2008uncoupling} (right). (B) Idealised representation of the expression domain of \textit{Ptc1} in panel A. Areas with consistently high expression in all images are shown in black, those with low expression in white. Gray areas indicate areas with ambigious expression patterns. In the cost function, the simulation output is penalized if the concentration within the black regions is less than 70\% of the maximal concentration and above 30\% in the white areas. The gray area is not considered when calculating the cost.}
\label{fig:cost}
\end{figure}

The first step of the parameter estimation process consists of defining the cost function. We develop a method for the case where the model is solved on a domain similar to to that of the shape of the tissue of interest from which our data is obtained, where the data to be reproduced is non-quantitative, noisy image-data. An example of such data is ISH data. ISH is a technique in which a complementary strand to the nucleic acid of interest, with a reporter molecule attached, is applied to the tissue of interest \cite{hargrave2006situ, mcfadden1995situ}. Upon addition of a substrate, a chemical reaction catalysed by the reporter molecule takes place which reveals the location of a specific segment of nucleic acid, for instance DNA or RNA. Variations in the experimental protocol, for example leaving the probe to bind for a different amount of time, washing for a longer period, or, in particular, the length of time of the chemical reaction, result in differences in the data obtained. In addition, non-specific binding of the probe leads to background signal. The result of this is non-quantitative and noisy image data, wherein the relationship between the intensity of the staining and the concentration of the nucleic acid of interest is non-linear, see Figure \ref{fig:cost}A. 

From the image data we determine a reference image, the shape of which is identical to the domain the model is solved on. We identify three separate areas in the data: the area where the staining is consistently darkest in all the data sets, i.e. the concentration is highest, the area where staining is consistently low across all data sets and the remaining area, where the staining is neither high nor low or varies between data sets. These three areas are marked on the reference image (the high concentration area in black, the low in white and the rest in gray), see Figure \ref{fig:cost}B. As the data is noisy and non-quantitative the gradient of gene expression can vary between images of the same gene (Figure \ref{fig:cost}A). However, usually it is clear which areas of the tissue of interest have reproducibly higher levels of gene expression and which have lower levels of expression. Therefore, we want our model to recapitulate those areas of high and low gene expression, rather than a set gradient, which can vary between images.

We then quantify the difference between the output of the model and the reference image. The reference image is grayscale i.e. black and white. White corresponds to a maximum pixel value of 255/255, and black corresponds to a minimum pixel value of 0/255. Shades of gray take on pixel values within this range. In order to compare the output of our model to the reference image we normalize the output of each component in the model we wish to compare with respect to its maximum concentration and plot the result in grayscale, with black (pixel value 0/255) representing the maximum concentration and white (pixel value 255/255) zero concentration. We now have two grayscale images that can be compared. 

We define the cost function as follows. Areas corresponding to the white area in the reference image are penalized linearly if their pixel value is smaller than 179/255 i.e. if the concentration is more than 30\% of the maximal value. This means that the penalty is zero if the pixel value, $v$, is within the range of 179 to 255 and otherwise equal to $179-v$. Similarly areas corresponding to the black area in the reference image are penalized linearly if their pixel value is larger than 76/255 i.e. the concentration is less than 70\% of the maximal value. Areas corresponding to the gray areas in the reference image are not included in the calculation. The cost function is then the squared sum of errors across all pixels across all images (one image for each variable we wish to include in the cost function), see Listing \ref{1st:label}.

By splitting the reference image into areas of high, mid and low concentration and normalizing with respect to the maximum concentration we remove the impact of noise from the data on the cost function and the problem of comparing non-quantitative data to the quantitative output of the model. Other thresholds are, of course, possible as well.

\begin{lstlisting}[frame=single, caption = {Cost Function}, label={1st:label}, language = C, basicstyle = \footnotesize, mathescape=true]  % Start your code-block

for images i
	for pixel indices p in image i
		if referencePicture(p,i)$\leq$76
			/* 70% concentration = 76 gray value */
			scorePic(p,i) = max(testPicture(p,i)-76,0)
		else if referencePicture(i,j)$\geq$179
			/* 30% concentration = 179 gray value */
			scorePic(p,i) = max(255-testPicture(p,i)-76,0)
		else
			scorePic(p,i) = 0
		end
	end
end
/* score = sum of squared entries of scorePic */
score = $\sum_{i,p}\text{scorePic(p,i)}^2$
\end{lstlisting}

\subsection{Parameter Estimation}
We use a Gaussian process to estimate the response surface with respect to our parameters. A Gaussian process is defined as a collection of possibly infinite random variables such that any finite collection of them has a multivariate normal distribution. A Gaussian process is fully defined by its mean function, $\mu(\bf{x},\boldsymbol{\theta})$ and covariance kernel function $k(\bf{x},\bf{x'};\boldsymbol{\theta})$.  In this case $\bf{x}, \bf{x'}\in $ $X$, where $X$ is the parameter space. The kernel parameters $\boldsymbol{\theta}$ are usually dependent on a characteristic length scale. The characteristic length scale defines roughly how far away two points have to be in order to be almost uncorrelated, see \cite{rasmussen2004gaussian} for an overview of Gaussian processes. With the mean and covariance function defined, one can then calculate the expected value of the response surface, which we call $f$, at a set point, given the value of $f$ at other points. It is also possible to calculate the standard deviation of $f$ at that point, which tells us how likely it is that the value of $f$ at that point will be close to the expected value calculated. 

We use the \verb|bayesopt| algorithm in MATLAB \cite{matlab} for our parameter estimation process which makes use of Gaussian processes to predict the response surface. It uses the automatic relevance determination (ARD) Matern 5/2 kernel. This is defined as
\begin{equation*}
k({\bf{x}},{\bf{x'}} | {\boldsymbol{\theta}}) = {\sigma_f}^2 \bigg(1+ \sqrt{5} r + \frac{5}{3}r^2 \bigg) \text{exp}(-\sqrt{5} r),
\end{equation*}
where
\begin{equation*}
r = \sqrt{\sum_{m=1}^{d}\frac{(x_{m}-x_{m}')^2}{{\sigma_m}^2}}.
\end{equation*}
Here, $\boldsymbol{\theta}$ is the vector of the kernel parameters. In this case
\begin{align*} 
\theta_m &= \log \sigma_m, \text{for} \ m = 1, 2, ...d \  \text{and} \\
\theta_{d+1} &= \log \sigma_f,
\end{align*}
where $d$ is the number of parameters that are being estimated, $\sigma_f$ is the standard deviation of the response surface and the$\  \sigma_m$ are the characteristic length scales for each parameter.
\verb|Bayesopt| estimates the parameters $\boldsymbol{\theta}$ of the covariance function from known values of the response surface. It then uses Gaussian process regression to predict the shape of the response surface and selects which point to sample at next. The theoretical convergence properties of this approach are analysed in \cite{bull2011convergence} and \cite{vazquez2010convergence}.

As the model to be solved is assumed to be computationally expensive, there is a limit to the number of feasible iterations. We therefore set a limit to the number of iterations we allow for the parameter estimation procedure. This number will differ depending on the model e.g. a model with a high number of parameters to be estimated or a highly irregular response surface will generally require more iterations for the identification of a suitable parameter set. A model that has a very high computational cost will have a lower number of feasible iterations than a model whose computational cost is less prohibitive. The set number of iterations is a trade-off between the accuracy of the parameter estimation process and the computational cost of the process. 

\begin{figure} [h]
\centering
\includegraphics[width=0.6\textwidth]{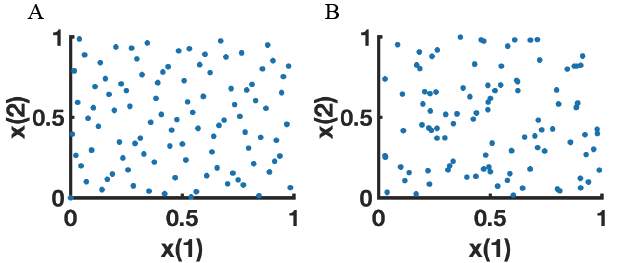}
\caption{Comparison of 100 points in 2D drawn from a Halton sequence and randomly (A) 100 points plotted from a Halton sequence. (B) 100 points plotted randomly. The points in panel A are more evenly distributed than in panel B. }
\label{fig:halton}
\end{figure}

For our model we limit the number of iterations to 400. Estimating a response surface is based on the knowledge of the values of the cost function at the parameter values sampled thus far. Therefore, we first run the model at parameter values spaced evenly in the sample space to gain a rough estimate of the shape of the response surface. We use a Halton sequence \cite{morokoff1994quasi} to generate 100 points that are spaced evenly in the parameter space. To generate these points we use the MATLAB function \verb|haltonset|. We choose to use a Halton sequence rather than selecting points randomly as a Halton sequence distributes points consistently in space (unlike a sequence of points generated randomly) \cite{kocis1997computational}, see Figure \ref{fig:halton}. See \cite{morokoff1994quasi} for details on generating Halton sequences. We input the parameter values at which we have already evaluated the model with the corresponding value of the cost function into the \verb|bayesopt| function as initial (already known) values. We allow the algorithm to run for a further 300 iterations (resulting in 400 iterations in total) and then select the parameter with the lowest corresponding cost function value that has been sampled by the algorithm as our optimal parameter.

\section{Results}

\subsection{Estimating the parameters of a test function}

\begin{figure} [h]
\centering
\includegraphics[width=0.5\textwidth]{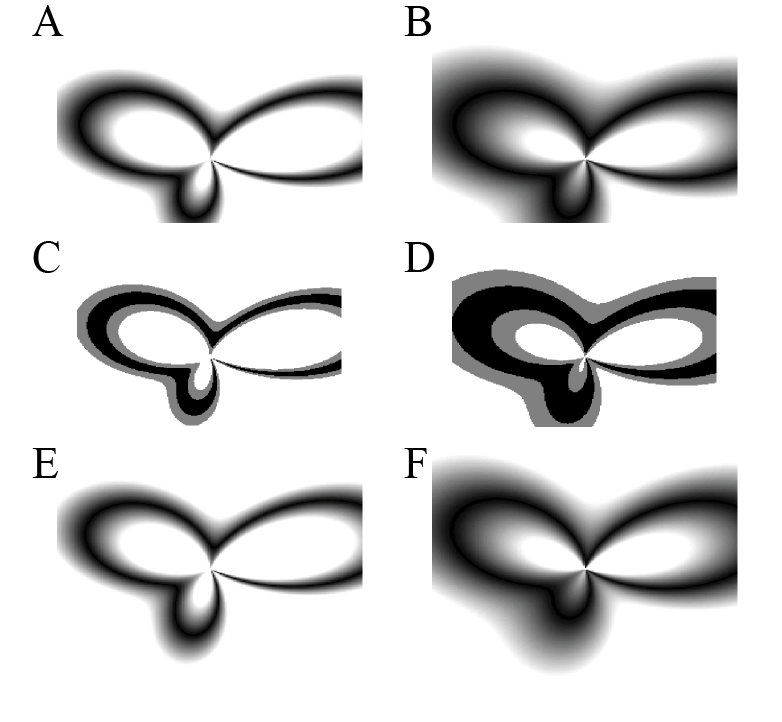}
\caption{Images generated from a parametric function with 7 parameters. (A, B) Original Images, (B) represents an image with more noise ($c=2$). (C, D): High (black) and low (white) concentration areas. Gray area is not included in cost function. (E, F): Example of reconstructed images from optimum parameter after optimization.}
\label{fig:parametric}
\end{figure}

To test our parameter estimation procedure we used it to  estimate the coefficients of the function
\[
f({\bf x},\boldsymbol{\theta})=\max\big\{0,1-c\big|\|{\bf x}\|_2-p({\arg{\bf x},{\boldsymbol{ \theta}}})\big|\big\},\quad c>0,
\]
where
\[
p(\alpha,\boldsymbol{\theta}) = \theta_0+\sum_{k=1}^{M-1}
+\theta_{2k}\cos(k\alpha)+\theta_{2k-1}\sin(k\alpha)+\theta_{2M-1}\cos(M\alpha), \quad  0 \leq \alpha < 2\pi
\]
is a trigonometric polynomial and $\alpha = \text{arg}\ \bf{x}$. We set the parameter $c$ to 4. Thus, the parameter estimation amounts in this case to estimating the \(2M\) coefficients \(\boldsymbol{\theta}\) of \(p(\alpha,\boldsymbol{\theta})\)
from images. We set $M=4$ and fix $\theta_2=0$ (excluding it from the parameter estimation procedure). We then try to estimate the remaining 7 parameters $\boldsymbol{\rho} = [\theta_0, \theta_1, \theta_3, \theta_4, \theta_5, \theta_6, \theta_7]$. We ran the parameter optimization procedure with a total of 400 iterations, of which the first 100 were initial Halon points. The range of the parameter space we sampled across was [0.01,1] for each parameter with the Halton set distributed across the log of the parameter space. This means that, for all parameters, the number of Halton points present in the range [0.01,0.1] should be approximately equal to the number present in the range [0.1,1]. The parameter estimation procedure was also performed on the log of the parameter space as we assumed changing a parameter from 0.01 to 0.1 would have a similar effect to that of changing a parameter from 0.1 to 1. We found that for 7 parameters we were able to successfully reconstruct the image, with the algorithm returning an optimum parameter set close to the original one, see Table \ref{table:parametric_results}. We then tested the algorithm on a more blurred image by changing the value of $c$ to 2 (corresponding to more ambiguous or noisy image data) and the algorithm was still able to reconstruct the image. See Figure \ref{fig:parametric}. To examine the effect of parameters that have little impact on the cost function we added dummy parameters into the algorithm that had no effect on the images. These were input into \verb|bayesopt| as parameters to optimize but were not inputs to the parametric function. The algorithm was still able to find a value close to the original parameter set for the 7 relevant parameters even with 10 dummy parameters added, see bottom row of Table \ref{table:parametric_results}.

\begin{figure} [h]
\centering
\includegraphics[width=0.8\textwidth]{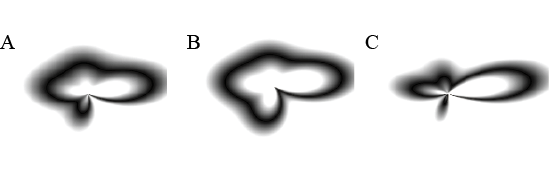}
\caption{Visualizations of parametric function for different parameter sets. (A) $\boldsymbol{\rho} = [0.3\  0.2 \ 0.1 \ 0.1\  0.1\  0.1\  0.1]$, (B) $\boldsymbol{\rho} = [0.5\  0.1 \ 0.1 \ 0.1\  0.1\  0.1 \ 0.1]$, (C) $\boldsymbol{\rho} = [0.1 \ 0.3 \ 0.2 \ 0.2 \ 0.2\  0.2 \ 0.2]$.}
\label{fig:examples}
\end{figure}

\begin{table}
\begin{center}
\caption{Comparison of original parameters (top) of the parametric function and optimum parameters found (sample size = 10, centre) and with dummy parameters added (single run, bottom). As the algorithm is not deterministic, distinct optimal values can emerge in independent runs. }
\begin{tabular}{ l  c  c  c  c  c  r }
\toprule			
 $\theta_0$ & $\theta_1$ & $\theta_3$ & $\theta_4$ & $\theta_5$ & $\theta_6$ & $\theta_7$ \\ \hline
 0.5 & 0.2 & 0.1 & 0.3 & 0.3& 0.2 & 0.1\\ 
 0.5 $\pm 0.01$ & 0.19 $\pm$0.02 & 0.08 $\pm$ 0.02 & 0.30 $\pm$ 0.02 & 0.30 $\pm$ 0.02 & 0.19 $\pm$ 0.01& 0.08 $\pm$ 0.02\\
 0.45 & 0.19 & 0.05 & 0.31 & 0.34 & 0.19 & 0.13\\
\bottomrule
\label{table:parametric_results}
\end{tabular}
\end{center}
\end{table}

\begin{figure} [h]
\includegraphics[width=0.5\textwidth]{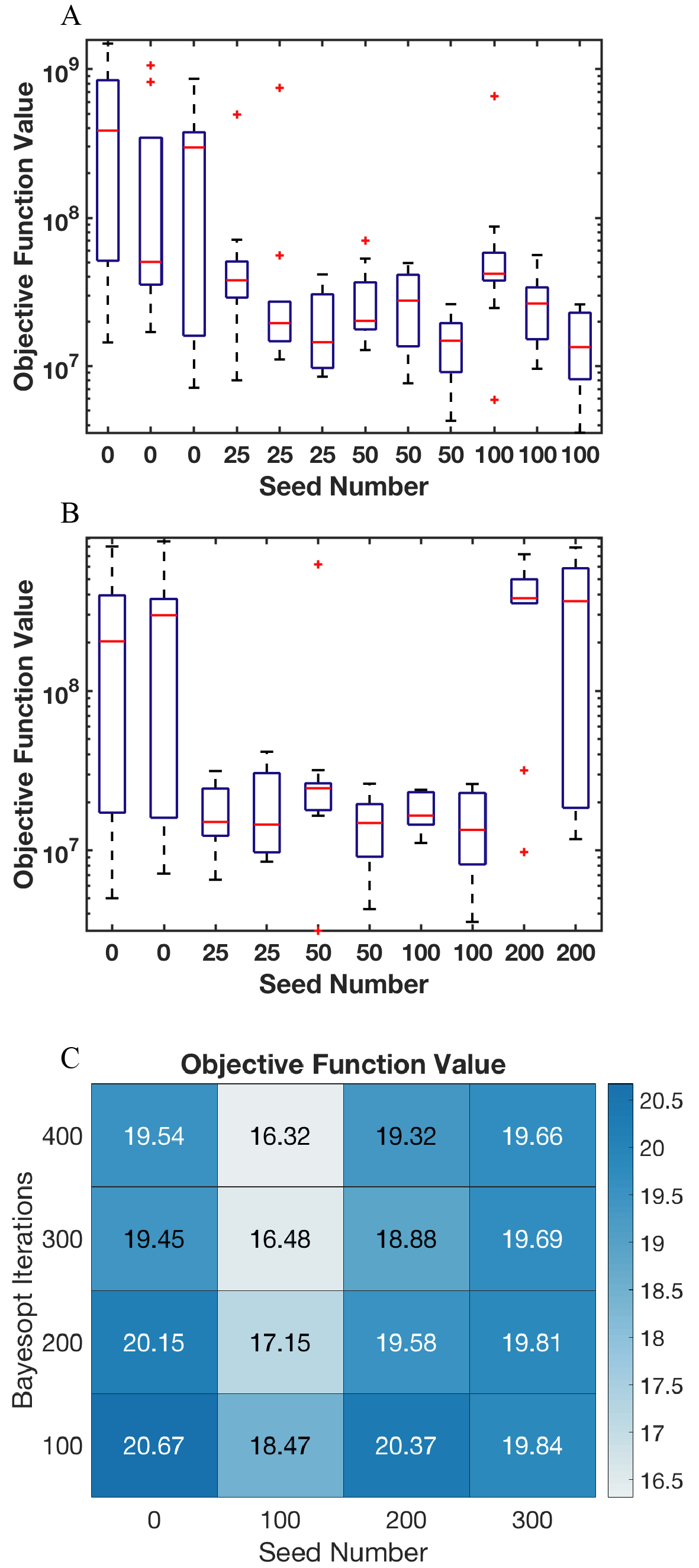}
\caption{Results of the parameter estimation procedure with different settings (A) Minimum objective function value found for different number of initial Halton points with 200 iterations (left boxplot), 300 iterations (middle boxplot), 400 iterations (right boxplot). (B) Minimum objective function value found for different number of initial Halton points with 400 iterations. GPActiveSetSize = 300 (left boxplot), GPActiveSetSize = 400 (right boxplot). Sample size = 10. (C) Mean minimum objective function value found for different number of initial Halton points and bayesopt iterations. Sample size = 10 for total iterations $\leq$ 400. Sample size = 5 for total iterations $>$ 400. Values are log scale.}
\label{fig:boxplot}
\end{figure}

We then looked at how decreasing the number of total iterations and changing the number of initial Halton points (seed number) affected the optimization procedure. We ran the optimization procedure with 200, 300 and 400 total iterations, with 0, 25, 50 and 100 initial Halton points. We found that simulations run with no initial Halton points performed considerable worse and those with 200 and 300 iterations did not perform as well as those run with 400 iterations. The results are shown in Figure \ref{fig:boxplot}A.

We further examined how varying the number of Halton points and the number of points \verb|bayesopt| used to predict the response surface affected the optimization procedure for a total of 400 iterations. The active set is defined as the set of points used to predict the response surface. The size of this set can be modified by changing the variable GPActiveSetSize in \verb|bayesopt|. For example, if \verb|bayesopt| has sampled at 350 points but the active set size is 300 then 300 points out of the 350 are selected uniformly at random without replacement to estimate the response surface. (The previous simulations had been run with the default GPActiveSetSize=300.) We tested each of these points with an active set size of 300 and an active set size of 400. The results are shown in Figure \ref{fig:boxplot}B. We found that the lowest median value was for 100 Halton points with GPActiveSetSize = 400. However, simulations run with 25 and 50 Halton points also performed well, whereas those with 200 and 0 Halton points performed considerably worse. This indicates that, for a set number of iterations, an initial set of Halton points can greatly improve the parameter estimation procedure, but too many prevent the algorithm from finding the optimum.

Finally, we looked at how varying the number of Halton points and the number of iterations performed after these initial simulations affected the parameter estimation procedure. We call the latter the number of bayesopt iterations. The total number of iterations is then the sum of the number of initial Halton points and the number of bayesopt iterations. We varied the seed number (number of initial Halton points) in increments of 100, from 0 to 300. We varied the number of bayesopt iterations from 100 to 400, also in increments of 100. The results are shown in Figure \ref{fig:boxplot}C. As expected, with the number of Halton points fixed, the parameter estimation procedure generally performed better the higher the number of bayesopt iterations. Unexpectedly, the converse does not appear to be true. With a fixed number of bayesopt iterations, increasing the seed number does not lead to a smaller minimum objective value.  For the values sampled, the optimum seed number for a fixed number of iterations was 100. This optimum seed number likely depends on the number of parameters to be estimated, the range of the parameters and the number of bayesopt iterations and should ideally be determined before the parameter estimation procedure. This could be done by running the procedure on a surrogate model which has the same number of parameters to be estimated as the original problem but is cheaper to compute.

\subsection {Comparison with other algorithms}
We then tested how this procedure compared to two other commonly used optimization algorithms: simulated annealing and  genetic algorithm. The functions \verb|simulannealbnd| and \verb|ga| in MATLAB use simulated annealing and genetic algorithm respectively to search for a function minimum. We ran both algorithms on our test function to try and retrieve the original parameters with 400, 700 and 1000 iterations. The results are shown in Table \ref{table:comparison}. As can be seen, neither \verb|simulannealbnd| nor \verb|ga| succeeded in retrieving the original parameters of the test function, even when the number of iterations was increased to 1000. 

\begin{sidewaystable}
\begin{adjustbox}{max width=\textwidth,keepaspectratio}
\begin{tabular}{ l  c  c  c  c  c  c  c  c  c r }
\toprule			
algorithm & iterations & time ({\si{\second}})& $\theta_0$ & $\theta_1$ & $\theta_3$ & $\theta_4$ & $\theta_5$ & $\theta_6$ & $\theta_7$ & objective \\ \hline
 N/A & N/A & N/A & 0.5 & 0.2 & 0.1 & 0.3 & 0.3& 0.2 & 0.1\\ 
\verb|bayesopt| & 400 & 2215 $\pm 515$ &0.5 $\pm 0.01$ & 0.19 $\pm$0.02 & 0.08 $\pm$ 0.02 & 0.30 $\pm$ 0.02 & 0.30 $\pm$ 0.02 & 0.19 $\pm$ 0.01& 0.08 $\pm$ 0.02 & 14293003\\
\verb|simulannealbnd|& 400 & 602 $\pm$ 17 & 0.32 $\pm$  0.12 & 0.55 $\pm$ 0.23 & 0.38 $\pm$ 0.23 & 0.57 $\pm$ 0.19 & 0.38 $\pm$ 0.19 & 0.46 $\pm$ 0.2 & 0.28 $\pm$ 0.17 & 516660821\\
\verb|simulannealbnd|& 700 & 1047 $\pm$ 90 & 0.35 $\pm$ 0.13 & 0.54 $\pm$ 0.22 & 0.31 $\pm$ 0.19 & 0.56 $\pm$ 0.18 & 0.37 $\pm$ 0.2 & 0.44 $\pm$ 0.21 & 0.24 $\pm$ 0.17 & 390425288\\
\verb|simulannealbnd|& 1000 & 1448 $\pm$ 5 & 0.35 $\pm$  0.13 & 0.50 $\pm$ 0.23 & 0.17 $\pm$ 0.1 & 0.53 $\pm$ 0.18 & 0.23 $\pm$ 0.17 & 0.28 $\pm$ 0.12& 0.11 $\pm$ 0.12 & 254855944\\
\verb|ga|& 400 & 725 $\pm$ 13 & 0.55 $\pm$ 0.05 & 0.1 $\pm$ 0.11 & 0.02 $\pm$ 0.02 & 0.26 $\pm$ 0.08 & 0.32 $\pm$ 0.04 & 0.09 $\pm$ 0.02 & 0.08 $\pm$  0.05 & 564173347\\
\verb|ga|& 700 & 1152 $\pm$ 11 & 0.55 $\pm$  0.03 & 0.11 $\pm$ 0.02 & 0.05 $\pm$ 0.03& 0.25 $\pm$ 0.04 & 0.31 $\pm$ 0.04 & 0.13 $\pm$ 0.07& 0.07 $\pm$ 0.03 & 405229775\\
\verb|ga|& 1000 & 1615 $\pm$ 31 & 0.55 $\pm$ 0.06 & 0.13 $\pm$ 0.07 & 0.05 $\pm$ 0.05 & 0.26 $\pm$ 0.04 & 0.32 $\pm$ 0.03 & 0.12 $\pm$ 0.05 & 0.08 $\pm$ 0.04 & 383413571\\
\bottomrule
\label{table:comparison}
\end{tabular}
\end{adjustbox}
\caption{Comparison of global parameter estimation algorithms. Original parameters of the parametric function (top row) and optimum parameters found by different algorithms (sample size = 10). As the algorithms are not deterministic, distinct optimal values can emerge in independent runs. Objective is average minimum objective function value found. Simulation time of parametric function = 1.42 $\pm$ 0.05s (sample size = 10).}
\end{sidewaystable}

\subsection{Estimating Network Parameters}

\begin{figure} [h]
\centering
\includegraphics[width=0.4\textwidth]{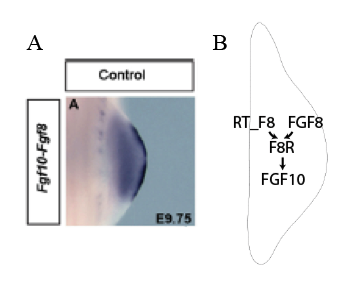}
\caption{Model network and gene expression (A) ISH data of \textit{Fgf10}, expressed in the mesenchyme (lighter staining), and \textit{Fgf8}, expressed in the distal AER (darker staining). Image is reproduced from \cite{sheth2013decoupling}. (B) Network scheme. FGF8 binds to its receptor and upregulates \textit{Fgf10} expression.}
\label{fig:network}
\end{figure}

We then tested our procedure on a more complex, biologically relevant model with 7 parameters. We based our model on the developing murine limb and fitted it to artificial ISH data of \textit{Fgf10} expression. During early limb development \textit{Fgf8} is expressed at the apical ectodermal ridge (AER), a structure at the distal end of the limb bud. It diffuses into the mesenchyme where it binds to its receptor to form a signaling complex (F8R). This complex upregulates the expression of another protein (\textit{Fgf10}), see Figure \ref{fig:network} for a network scheme and the corresponding ISH expression data. We solved our model on a realistic growing domain with domain size (initial size:160 \si{\micro\metre} from AER to flank and final size: 240 \si{\micro\metre} from AER to flank) and time (5 hours) corresponding to the development of the limb bud from somite stage 21 to somite stage 24. \textit{Fgf10} expression at the time point corresponding to somite stage 24 was used to compare against \textit{Fgf10} expression evaluated at the same time point from artificial ISH data (generated with parameter set $\boldsymbol{\theta_0}$).

The following reaction-advection-diffusion equations are used to represent the gene expression dynamics with the terms to be interpreted as follows. The variables $C_\text{F8}$, $C_\text{F10}$ and $C_\text{F8R}$ represent the concentration of FGF8, FGF10 and the FGF8-receptor-ligand signalling complex respectively. As the receptor is only present in the mesenchyme \cite{macarthur1995fgf, ornitz1996receptor}, we define the signalling complex $C_\text{F8R}$ to be present exclusively in this part of the computational domain. FGF10 and FGF8 are diffusible and can therefore be present in both the mesenchyme and AER domain. $p_\text{F10}$ and $p_\text{F8}$ are the production rates of FGF10 and FGF8, respectively. We set our initial values to zero. The initial expression of $C_\text{F10}$ is then also zero. As \textit{Fgf10} is expressed early in the limb bud, we add an additional production term to $C_\text{F10}$: $p_0\text{Fgf10}$. The indicator functions $\mathbbm{1}_\text{Mes}$ and $\mathbbm{1}_\text{AER}$ are used to show that \textit{Fgf10} and \textit{Fgf8} are expressed in the mesenchyme and AER respectively. The diffusion rate of FGF10, $D$, is assumed to be equal to that of FGF8 as they are both diffusible ligands. The diffusion rate of the signalling complex is denoted $D_r$. Similarly, $d$ is the degradation rate of FGF8 and FGF10, and $d_r$ that of the signalling complex. FGF8 binds to its receptor at a rate of $k_\text{on8}$ and the unbinding rate of the receptor-ligand complex is $k_\text{off8}$. As the FGF8 receptor is only present in the mesenchyme, binding and unbinding only occur in that part of the domain. The total receptor concentration is assumed to be in quasi-steady-state, we therefore use $RT_{F8}$ to denote the receptor concentration, so $(\text{RT}_\text{F8}-C_\text{F8R})$ is the concentration of free receptor. Assuming mass action kinetics, the binding and unbinding terms are $\mathbbm{1}_\text{Mes}k_\text{on8}C_\text{F8}(\text{RT}_\text{F8}-C_\text{F8R})$ and $\mathbbm{1}_{Mes}k_\text{off8}C_\text{F8R}$ respectively. Finally, we model $C_\text{F8R}$-induced expression by a Hill function $\frac{C_\text{F8R}^2}{C_\text{F8R}^2+K_{\text{F8R\_ F10}}^2}$ with Hill constant $K_{\text{F8R\_ F10}}^2$ and Hill coefficient 2. The velocity field of the tissue, due to growth, is denoted $\bold{u}$. 

\begin{align*}
\frac{\partial C_{\text{F10}}}{\partial t} + \nabla(\bold{u}C_{\text{F10}}) &= D \Laplace C_{\text{F10}}  + \mathbbm{1}_\text{Mes} \bigg(p_{\text{F10}}\frac{C_{\text{F8R}}^2}{C_{\text{F8R}}^2+K_{\text{F8R\_ F10}}^2}+ p_{0\text{Fgf10}}\bigg)-d C_{\text{F10}},\\
\\
\frac{\partial C_{\text{F8}}}{\partial t} + \nabla(\bold{u}C_{\text{F8}}) &= D \Laplace C_{\text{F8}}  + \mathbbm{1}_\text{AER}p_{\text{F8}}- d C_{\text{F8}}-  \mathbbm{1}_\text{Mes}k_\text{on8}C_{\text{F8}}(\text{RT}_{\text{F8}}-C_{\text{F8R}}) +\mathbbm{1}_{Mes}k_\text{off8}C_{\text{F8R}},\\
\\
\frac{\partial C_{\text{F8R}}}{\partial t} + \nabla(\bold{u}C_{\text{F8R}}) &=  D_{r} \Laplace C_{\text{F8R}}  + k_\text{on8}C_{\text{F8}}(\text{RT}_{\text{F8}}-C_{\text{F8R}})-(k_\text{off8}+d_{r})C_{\text{F8R}}.\\ 
\end{align*}

While the length and time scales of the developmental processes are typically well known, the absolute value of protein concentrations are largely unknown (due to ISH being non-quantitative). The standard procedure would be to non-dimensionalize the model to reduce the number of parameters. However, given the available information regarding the developmental length and time scales, we will retain dimensional length and time scales, and only eliminate two parameters by scaling the variables with respect to their particular production rates, i.e.

\begin{equation*}
C_{\text{F10}}=p_{\text{F10}}\widetilde{C_{\text{F10}}}, \ C_{\text{F8}}=p_{\text{F8}}\widetilde{C_{\text{F8}}},\ C_{\text{F8R}}=p_{\text{F8}}\widetilde{C_{\text{F8R}}}, \ \text{and} \ RT_{\text{F8}}=p_{\text{F8}}\widetilde{\text{RT}_\text{F8R}}.
\end{equation*}

We get 

\begin{align*}
\frac{\partial \widetilde{C_\text{F10}}}{\partial t} + \nabla(\bold{u}\widetilde{C_{\text{F10}}})  &= D \Laplace \widetilde{C_\text{F10}}  + \mathbbm{1}_\text{Mes}\bigg(\frac{ \widetilde{C_\text{F8R}^2}}{\widetilde{C_\text{F8R}^2}+\frac{K_\text{F8R\_ F10}^2}{{p_\text{F8}}^2}}+  \frac{p_{0\text{Fgf10}}}{p_\text{F10}}\bigg)- d\widetilde{C_\text{F10}},\\
\\
\frac{\partial \widetilde{C_\text{F8}}}{\partial t} + \nabla(\bold{u}\widetilde{C_{\text{F8}}}) &= D \Laplace \widetilde{C_\text{F8}}  +\mathbbm{1}_\text{AER} - d\widetilde{C_\text{F8}} +  \mathbbm{1}_\text{Mes}\big(k_{\text{off8}}\widetilde{C_\text{F8R}} -k_{\text{on8}}p_{\text{F8}}\widetilde{C_\text{F8}}(\widetilde{\text{RT}_\text{F8}}-\widetilde{C_\text{F8R}})\big), \\
\\
\frac{\partial \widetilde{C_\text{F8R}}}{\partial t} + \nabla(\bold{u}\widetilde{C_{\text{F8R}}}) &=  D \Laplace \widetilde{C_\text{F8R}}  +k_\text{on8}p_{\text{F8}}  \widetilde{C_\text{F8}}(\widetilde{\text{RT}_{\text{F8}}}-\widetilde{C_\text{F8R}}) -(k_\text{off8}+d_r)\widetilde{C_\text{F8R}}. \\
\end{align*}

We assume $k_\text{off8} \ll d_r, \ k_{\text{on8}}p_\text{F8}$. Then, grouping parameters by settting: $\widetilde{K_{F8R\_ F10}^2} = K_\text{F8R\_ F10}^2/{p_\text{F8}}^2, \ \widetilde{k_\text{on8}}= k_\text{on8}p_\text{F8}$ and $\widetilde{p_{0\text{Fgf10}}}=p_{0\text{Fgf10}}/p_{\text{F10}}$ and dropping the tildes for convenience we obtain the final set of equations:

\begin{align*}
\frac{\partial {C_\text{F10}}}{\partial t} + \nabla(\bold{u}C_{\text{F10}})  &= D \Laplace {C_\text{F10}}  + \mathbbm{1}_\text{Mes}\bigg(\frac{ {C_\text{F8R}^2}}{{C_\text{F8R}^2}+ {K_{\text{F8R\_ F10}}^2}}+p_{0\text{Fgf10}}\bigg)- d{C_\text{F10}},\\
\\
\frac{\partial {C_\text{F8}}}{\partial t} + \nabla(\bold{u}C_{\text{F8}})  &= D \Laplace {C_\text{F8}}  +\mathbbm{1}_\text{AER}- d{C_\text{F8}}+\mathbbm{1}_\text{Mes}\big( -{k_\text{on8}}{C_\text{F8}}({\text{RT}_{\text{F8}}}-{C_\text{F8R}})\big), \\
\\
\frac{\partial {C_\text{F8R}}}{\partial t} + \nabla(\bold{u}C_{\text{F8R}}) &=  D_r \Laplace {C_\text{F8R}}  +{k_\text{on8}}{C_\text{F8}}({\text{RT}_{\text{F8}}}-{C_\text{F8R}}) -d_r{C_\text{F8R}}. \\
\end{align*}

\begin{figure} [h]
\centering
\includegraphics[width=0.6\textwidth]{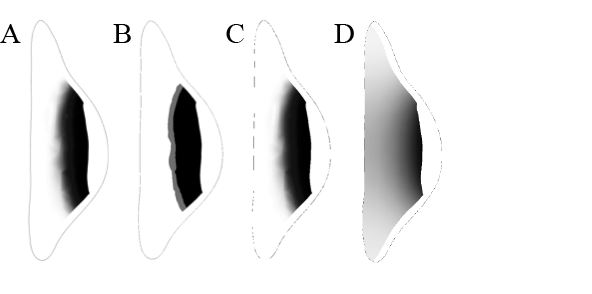}
\caption{Image comparison (A) \textit{Fgf10} expression for initial parameter set ($\boldsymbol{\theta_0}$).
(B) Reference image from initial parameter set with areas of high (black), low (white), and intermediate (gray) expression levels.
(C) Image of \textit{Fgf10} expression corresponding to the optimum parameter set. (D) Image of \textit{Fgf10} expression corresponding to the optimum parameter set found with the simplified 2-equations model.}
\label{fig:results}
\end{figure}

The diffusion rate of the ligand, $D$, is fixed as $100$ times higher than the diffusion rate of the receptor $D_r$. The artificial ISH data of \textit{Fgf10} corresponds to the production rate term of FGF10, i.e. $\mathbbm{1}_\text{Mes}\bigg(\frac{ {F8R^2}}{{F8R^2}+ {K_{F8R\_ F10}^2}}\bigg)$. The 7 network parameters to be estimated are then as follows: the diffusion coefficient for FGF8 ($D$), the ligand degradation rate ($d$), the signaling complex degradation rate ($d_r$), the binding rate between FGF8 and its receptor ($k_\text{on8}$), the total amount of receptor ($RT_{F8}$) and the Hill coefficient for enhanced production of \textit{Fgf10} (${K_{F8R\_ F10}^2}$). The seventh parameter is the initial FGF10 production rate ($p_0\text{Fgf10}$). We took an initial parameter set $\boldsymbol{\theta_0}$ (see Table \ref{table:network_parameters}) and used it to generate a reference image for expression of \textit{Fgf10} at the time point corresponding to somite stage 24: see Figure \ref{fig:results}. We then ran our parameter estimation process, using the \textit{in silico} data to calculate the cost function. We ran the parameter estimation procedure with 100 Halton points and 400 iterations in total with an active set size of 400. The optimum parameter set found by the algorithm resulted in an expression pattern that was almost identical to the input image, see Figure \ref{fig:results}. However, the optimum parameter value that was returned varied significantly from our initial parameter set $\boldsymbol{\theta_0}$, with only the degradation rate of the signaling complex within 10\% of the original parameter value, see Table \ref{table:network_parameters}.

\begin{sidewaystable}
\caption{Comparison of original parameters (top) of the network and optimum parameters found (bottom)}
\small
\begin{tabular}{ l  c  c  c  c  c  r }
\toprule			
 $D=\theta_1$ (\si{\micro\metre^2\hour^{-1}})& $d=\theta_2$ ({\si{\hour^{-1}}}) & $d_r=\theta_3$ ({\si{\hour^{-1}}})& $k_\text{on8}=\theta_4$ ({\si{\hour^{-2}}})& $RT_{F8}=\theta_5$ ({\si{\hour}})& ${K_{F8R\_ F10}}=\theta_6$ ({\si{\hour}})& $p_0\text{FGF10}=\theta_7$ \\ \hline
16000 & 0.1 & 0.3 & 60 & 5 & 0.15 & 0.1\\ 
 16344 & 0.216 & 0.627 & 87.7 & 2.38 & 0.101& 0.0146\\
\bottomrule
\label{table:network_parameters}
\end{tabular}
\end{sidewaystable}

In order to understand why the algorithm failed to retrieve the original parameter values, although the image representing \textit{Fgf10} expression at the optimum value found was almost identical to the image representing \textit{Fgf10} expression for the initial parameter set $\boldsymbol{\theta_0}$, we ran sensitivity analyses for the model. We also ran sensitivity analyses for the parametric function, where the parameter estimation process had accurately retrieved the original image as well as the original parameter set.  

\begin{figure} [h]
\centering
\includegraphics[width=1\textwidth]{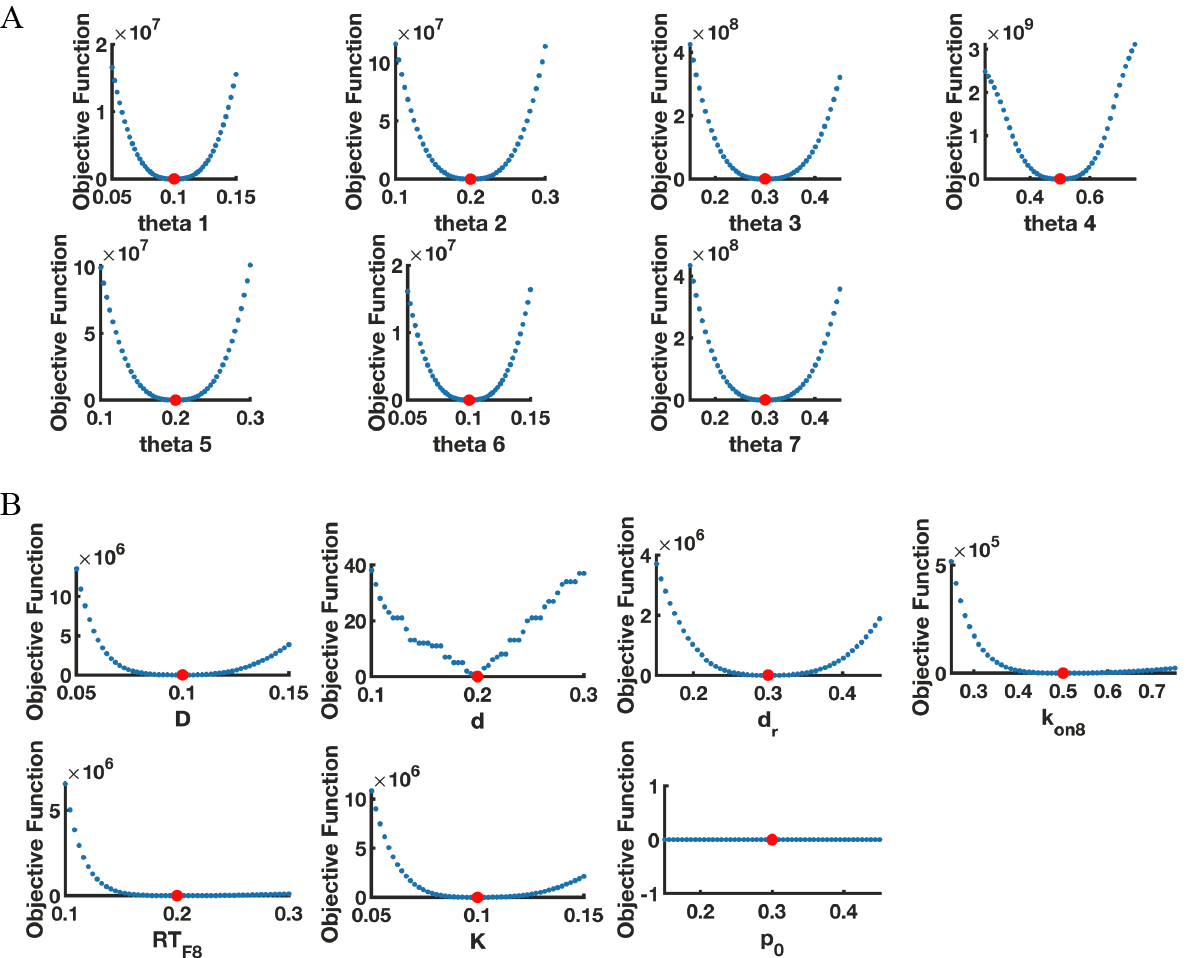}
\caption{Sensitivity analysis about the optimum: (A) parametric function and (B) limb model parameters. }
\label{fig:sensitivities}
\end{figure}

The results of the single variable sensitivity analysis can be seen in Figure \ref{fig:sensitivities}. The response curve for the parametric function is convex about the optimum and continuous. For the parameters in the limb network the response curve for $\theta_2$ is discontinuous. For $\theta_7$ it is flat, implying that this parameter has no impact on the cost function. For parameters $\theta_5$, $\theta_4$ and $\theta_1$ the response curve appears almost flat for a wide subset of parameters about or on one side of the optimum. This could be a reason as to why the optimum parameters found for parameters $\theta_4$, $\theta_7$, $\theta_2$ and $\theta_5$ are different to the initial $\boldsymbol{\theta_0}$. 

\begin{figure} [h]
\centering
\includegraphics[width=1\textwidth]{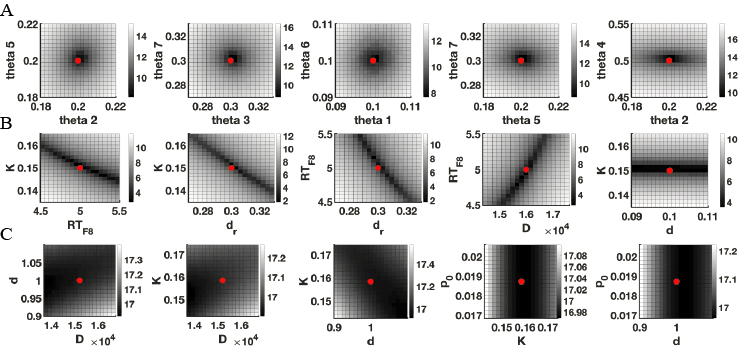}
\caption{Multivariate sensitivity analysis about the optimum for three different models.  (A) Parametric function,  (B) limb model with seven parameters, (C) simplified limb model with four parameters. Values are log scale.}
\label{fig:multivariate}
\end{figure}

The results of the multivariate sensitivity analysis, for selected parameters, can be seen in Figure \ref{fig:multivariate}B. The impact of the parameters of the parametric function on the cost function is independent. On the other hand, parameter dependencies can be seen for the limb network model. For example, if we look at parameters $\theta_3$ and $\theta_5$ ($d_r \ \text{and} \ RT_{F8}$) we see that if $\theta_3$ is fixed at a higher value, the minimum value of the cost function is obtained by decreasing $\theta_5$. If we look at the optimum parameter set that was returned and the original values of $\theta_3$ and $\theta_5$ we can see this. For the optimum value returned, $\theta_3$ is higher than the original value and $\theta_5$ is correspondingly lower. Similarly, for parameters $\theta_3$ and $\theta_6$ ($d_r \ \text{and} \  K_\text{F8R\_F10}$), for the optimum value returned, $\theta_3$ is higher than the original value and $\theta_6$ is correspondingly lower.

We further looked at whether it was possible to simplify the model. To this end, we implemented a direct activation of $C_\text{F10}$ by $C_\text{F8}$, thereby removing $C_\text{F8R}$ from the model and simplifying the system. This resulted in the following equations:

\begin{align*}
\frac{\partial C_\text{F10}}{\partial t}  + \nabla(\bold{u}C_{\text{F10}}) &= D \Laplace C_\text{F10}  + \mathbbm{1}_\text{Mes}\bigg(\frac{ {C_\text{F8}^2}}{{C_\text{F8}^2}+ {K^2}} + p_0\text{Fgf10}\bigg)- dC_\text{F10},\\
\\
\frac{\partial C_\text{F8}}{\partial t}  + \nabla(\bold{u}C_{\text{F8}}) &= D \Laplace C_\text{F8}  +\mathbbm{1}_\text{AER}- dC_\text{F8} \\
\end{align*}

with the 4 parameters: $D, \ d, \ K\  \text{and} \  p_0\text{FGF10}$. We then ran the parameter estimation procedure, using the same reference image as for the original model, and performed a multivariate sensitivity analysis about the optimum found. The optimum parameters found are shown in table \ref{table:simplified_parameters} and the sensitivity analysis in Figure \ref{fig:multivariate}C. Interestingly, the optimum parameters found were close to the corresponding original parameters of the original network. However, the image of \textit{Fgf10} expression was considerably different to that of the reference image, see Figure \ref{fig:results}D. This is likely due to the difference in diffusion rate between $F8$ and $F8R$, leading to a sharper boundary in the reference image, resulting in a qualitatively different result when $F8R$ is removed from the equations. From the sensitivity analysis it can be seen that parameter dependencies are still present, even in this simplified model, for instance between $D$ and $d$ and $D$ and $K$.

\begin{table}
\begin{center}
\caption{Optimum parameters found for simplified model}
\begin{tabular}{ l  c  c  r }
\toprule			
 $D=\theta_1$ ({\si{\micro\metre^2\hour^{-1}}})& $d=\theta_2$ ({\si{\hour^{-1}}}) & $ K_{F8\_F10}=\theta_3$  ({\si{\hour}})& $p_0\text{FGF10}=\theta_4$ \\ \hline
15212  & 1 & 0.16 & 0.019\\ 
\bottomrule
\label{table:simplified_parameters}
\end{tabular}
\end{center}
\end{table}

\clearpage

\section{Discussion}
The importance of global optimization methods for parameter estimation in biological modelling is well recognized (due to the complexity of the cost function with respect to the parameters) and numerous methods have been developed for this purpose. Most require a large number of model simulations to obtain good results. This is not feasible if the model is computationally costly, thereby prohibiting multiple iterations. In this paper, we proposed a heuristic global parameter estimation procedure for image-based modelling where the image data is noisy and non-quantitative and the model is computationally expensive to solve. To this end, we described a cost function, which measures the distance between the current and desired output of the model, and a procedure, using Gaussian process learning, to subsequently perform the parameter estimation. 

Using an image-generating parametric function to test the process we showed that, for seven parameters, this method retrieves the original parameters (and therefore the original image) when the parameters are independent. We then tested the procedure on a biological network with artificial data. As is often the case in biological models \cite{gutenkunst2007universally}, even with simple networks, there was a high level of parameter dependence in the network model. Applying the parameter estimation procedure resulted in a parameter set corresponding to an output image closely resembling the reference image. However, due to the parameter dependence, the optimal parameter set found differed considerably from that of the parameter set used to generate the reference image. 

There are several factors which influenced the parameter dependence in our model. For dynamical models, such as the limb model in our paper, a necessary condition to fix parameter information is the availability of temporal information. If the time scale is unknown, there will necessarily be parameter dependence as parameter sets that result in the same behaviour of the model, differing only in the rate of that behaviour, will be equally valid. In this case, parameter dependence due to lack of temporal information can be removed by non-dimensionalizing the equations with respect to time. In our limb model, we only evaluate the similarity of the model to the data at one time point (somite stage 24). This is one reason we have high parameter dependence: as parameter sets that fit the data equally well at this point, but result in different gene expression patterns at later points, have the same cost function value. Evaluating at additional time points, for instance at later somite stages, would decrease the parameter dependencies in our model, albeit at an increased computational cost due to increased simulation time of the model. In particular, this should decrease the dependencies between the degradation of the receptor $d_r$ and the diffusion coefficient $D$ and the degradation of the receptor $d_r$ and the Hill constant $K$. 

In addition, the only read-out we have is the production rate of FGF10 which we compare against artificial expression data. Including read-outs for other variables would decrease the parameter dependence, as well. For example, there is parameter dependence between $K$ and parameters that control the concentration of F8R e.g. $d_r, \ k_\text{on8}, \ RT_\text{F8}$ as the production rate of FGF10 can be modified by either changing the sensitivity of FGF10 to F8R (via $K$) or by changing the concentration of F8R. Both the sensitivity $K$ and the concentration of the FGF8-receptor complex are difficult to measure. But if data was available such that, for instance, the concentration of F8R could be added to the cost function, then the parameter dependence between $K$ and $\text{RT}_\text{{F8}}$ and $K$ and $d_r$ should decrease. The level of parameter dependence is contingent on the amount of data that is available to evaluate the goodness of fit of the model. In cases where the concentration of all the components in the model is known at several time points, the parameter dependence should be low. However, this is rarely the case in biological modelling, where either experiments can be expensive to run (prohibiting measurements at several time points) or substances can be difficult to measure (in which case the concentration of several components in the model will be unknown). 

Other than testing the validity of underlying assumptions of the biological system, the main purpose of computational modeling is to obtain predictions. Although additional experiments can be performed to constrain model parameters \cite{kreutz2009systems}, in most cases, the degree of measurement accuracy required to constrain predictions is extremely high and therefore impractical \cite{gutenkunst2007universally}. Therefore, the impact of parameter dependence on model behaviour should be examined. If the model behaviour is insensitive to certain parameter combinations, then parameter dependence does not rule out accurate predictions. In this case, the model output that determines model predictions is as insensitive to certain parameter combinations as the model output that is used to fit the model, and predictions can be made despite parameter non-identifiability \cite{gutenkunst2007universally}. Model predictions under different sets of feasible parameters can be analysed (the ensemble method) \cite{battogtokh2002ensemble, brown2004statistical}. For example, uncertainty estimates for the model output can be calculated by generating an ensemble of feasible parameters and calculating the mean and standard deviation of model output. This approach is illustrated in \cite{brown2004statistical}, where experimental verification confirms model predictions, despite large parameter uncertainties in the model. However, in cases where the model is computationally expensive, determining and simulating the sets of feasible parameters could have a prohibitive computational cost. 

When considering which global optimization method to use, an important consideration is the computational cost of the model. The simulation time of our limb model was approximately 40 seconds, running 400 simulations therefore took approximately 4.5 hours. The Gaussian process regression algorithm has a time complexity of $O(n^3)$, where $n$ is the number of training data. Hence, if the simulation is relatively cheap, the ratio of computational time of the algorithm to the computational time of the simulations can be high. In this case, global optimization algorithms with lower computational complexity, such as genetic algorithms, can perform better. However, as mentioned in the introduction, these algorithms often require a high number of iterations to find an optimum. 

We used the function \verb|ga| in the MATLAB optimization toolbox to test how genetic algorithms would perform on our parametric function. When the function was run with a maximum of 400 iterations the optimum found by \verb|ga| was far from the original input values. The image obtained with the optimum found was also significantly different from the original image. With 700 iterations \verb|ga| again failed to retrieve the original parameters, however, the image obtained by the optimum found was similar to the original image. Using the function \verb|ga| with 700 iterations of our parametric function took, on average, 1152 seconds. As one model simulation, on average, of our parametric function was 1.42s this meant that 994s could be accounted for from just running the simulations. Therefore, \verb|ga| spent approximately 158s on operations other than running the simulation (a relatively small fraction of the total running time). In contrast, running \verb|bayesopt| with 400 iterations of our parametric function took 2215s. This meant that 568s could be accounted for from just running the simulations, and then \verb|bayesopt| spent approximately 1647s on operations other than running the simulation (a considerably larger fraction of the total running time). Consequently, for our parametric function, running 700 iterations of \verb|ga| is less costly than 400 iterations of \verb|bayesopt|. In contrast, the simulation time for the limb model is 40 seconds. Running \verb|bayesopt| with 400 iterations then takes approximately $400 \times 40\si{s} + 1647\si{s}=17,647\si{s}$. In comparison, running \verb|ga| with 700 iterations would take a minimum of $700 \times 40\si{s} + 158\si{s}=28,158\si{s}$. Accordingly, the computational cost of the model determines which global optimization methods are feasible to employ.

In this paper, we restricted our models to those with seven parameters or less. To apply this method to estimate a larger number of parameters most likely a larger number of iterations would be required, and would need to be determined, along with an appropriate ratio of Halton points (initial seeds) to total iterations. The total number of iterations required could become prohibitively large for models with a large number of parameters and high computational complexity. A solution to this could be having a large number of initial seeds which are computed in parallel (if this is possible), thereby reducing the number of iterations that have to be performed in serial for the rest of the optimization procedure. However, as can be seen in Figure \ref{fig:boxplot}C, having a larger number of initial seeds does not necessarily lead to better results. It would therefore be useful to determine the optimum seed number (for the number of parameters to be estimated and the number of feasible bayesopt iterations) and then run these initial seeds in parallel. 

To calculate our cost function, we generated a reference image from the image data available. This reference image discriminates areas of high and low concentration, as well as the area that should be excluded from the cost function calculation. Depending on the features of the data that the model output should match, a different number of areas can be identified and penalised accordingly. For example, if a subset of the image should have a medium intensity, this area can also be identified and incorporated into the cost function, by penalising model outputs that result in concentrations that are too high or too low. What percentage of the maximum concentration is considered a high or low concentration depends on the model and data and the cost function can be implemented accordingly. The parameter estimation procedure can also be applied in cases where a portion, or all, of the data is quantitative. In this case, the cost function is simply modified by penalising the difference between the model output and the absolute concentrations expected rather than the maximum concentration output of the model. 

We expect that the method described in this paper can be used to perform global parameter estimation when model simulation is computationally costly and quantitative data is missing. The step of generating the reference image from the available data could be automised using image processing. This would facilitate an unbiased determination of the cost function from the data and the reference image could be updated as new data becomes available. The method could also be used as a model selection tool \cite{sivia2006data}, by treating model structure as a parameter, where different network topologies are compared to see which produces the best fit to the data.

\section{Software}
The parametric function was solved using MATLAB (version R2017b)  \cite{matlab}. The model of the developing limb was solved with COMSOL Multiphysics 5.1 \cite{comsolmultiphysics} and MATLAB Livelink (version R2017b) \cite{matlab} as described in \cite{germann2012simulating}.

\section{Acknowledgments}
We thank Harold G\'{o}mez and Marcelo Boareto for critically reading the manuscript, and Guido Sanguinetti for discussions. This work was supported by a SystemsX.ch iPhD grant (2014/240 to D.I.) and the Swiss National Science Foundation (SNF.170930 to D.I.).

\clearpage
\bibliographystyle{unsrt}
\bibliography{literature}
\end{document}